\documentclass[]{aastex7}
\newcommand\integ{INTEGRAL}
\newcommand\isgri{ISGRI}

\newcommand{\Tn}{\ensuremath{T_{90}}}
\newcommand{\Tbb}{\ensuremath{T_{\mathrm{BB}}}}

\newcommand{\Ep}{E$_{peak}$}
\newcommand{\sgros}{SGR 1806--20}
\newcommand{\sgrtf}{SGR 1935+2154}
\newcommand\onee{1E 1547--5408}
\def\rchisq {$\chi_{r} ^{2}$}
\def\flux {\mbox{erg cm$^{-2}$ s$^{-1}$}\ }
\def\fluence {\mbox{erg cm$^{-2}$}\ }

\usepackage{graphicx}
\usepackage{subcaption}
\usepackage{xcolor}

\begin{document}
\title{INTEGRAL IBIS catalog of magnetar bursts}

\author[0009-0001-3911-9266]{Dominik Patryk Pacholski} 
\affiliation{INAF, Istituto di Astrofisica Spaziale e Fisica Cosmica di Milano, via Corti 12, I-20133 Milano, Italy}
\affiliation{Universit\`a degli Studi di Milano Bicocca, Dipartimento di Fisica G. Occhialini, Piazza della Scienza 3, I-20126 Milano, Italy}
\email[show]{dominik.pacholski@inaf.it}

\author[0000-0003-3259-7801]{Sandro Mereghetti}
\affiliation{INAF, Istituto di Astrofisica Spaziale e Fisica Cosmica di Milano, via Corti 12, I-20133 Milano, Italy}
\email[show]{sandro.mereghetti@inaf.it}

\author[0000-0002-5711-9278]{Martin Topinka}
\affiliation{INAF, Osservatorio Astronomico di Cagliari, via della Scienza 5, I-09047  Selargius, Italy}
\email[show]{martin.topinka@inaf.it}

\begin{abstract}
One of the distinctive properties of magnetars, young neutron stars powered mainly by magnetic energy, is the emission of short   ($\lesssim$1 s) bursts of hard X-rays. Such bursts have been observed in nearly all the known magnetars, although at different and time-variable rates of occurrence. In the last two decades, the INTEGRAL satellite has extensively covered with good imaging capabilities the Galactic plane, where most magnetars reside.
We present the results of a comprehensive search for magnetar bursts in more than twenty years of archival data of the \integ\ IBIS instrument (15 keV -- 1 MeV).  
This led to the detection of 1349 bursts with 30-150 keV fluence in the $\sim2\times10^{-9} - 3\times10^{-6}$ erg cm$^{-2}$ range from 21 of the  34 examined magnetars and candidate magnetars with well known positions. The durations of the bursts, in terms of \Tn , follow a lognormal distribution centered at $\sim0.1$ s. 
Most of the detected bursts originated from three particularly active sources: \onee, \sgros, and \sgrtf. The integral  distributions of their burst fluences follow power laws with slopes $\beta$= 0.76$\pm$0.04, 0.95$\pm$0.06, and 0.92$\pm$0.10, respectively.
The burst spectra are generally well fit with an exponentially cut-off power law with    peak energy $E_{peak}$ in the range $\sim20-60$ keV for \sgros\  and \sgrtf , while the bursts of \onee\ are slightly harder ($E_{peak}\sim35-100$ keV).
A significant anti-correlation between $E_{peak}$ and fluence is found for \sgros , which provided the largest number of bursts among the sources of  our sample.
\end{abstract}

\section{Introduction}
Magnetars are young, isolated neutron stars with the most extreme magnetic fields, typically in the range of $10^{12}-10^{15}$ G \citep{1998Natur.393..235K,2015SSRv..191..315M, 2017ARA&A..55..261K}. They have spin periods of $\sim2-12$ s and relatively high spin-period derivatives of $10^{-11}$–$10^{-13}$ s s$^{-1}$. Unlike rotation-powered pulsars or neutron stars in accreting X-ray binaries, magnetars are believed to be powered by their strong internal magnetic fields \citep{1996ApJ...473..322T}. 

Since they are typically faint in their quiescent state, with luminosities of $10^{32}-10^{35}$ erg s$^{-1}$, many magnetars were first discovered through the detection of  short bursts, one of their most defining characteristics (hence the name of ``soft gamma-ray repeaters'' with which they were initially designated). These bursts, seen mainly in the hard X-ray,  are believed to originate in the magnetosphere, triggered by sudden crust failures caused by internal magnetic stresses \citep{1995MNRAS.275..255T,2001ApJ...561..980T} or by  rapid reconfigurations of the magnetic field \citep{2003MNRAS.339..623L,2010MNRAS.407.1926G,2013ApJ...774...92P}.

Magnetar bursts span a wide range of durations and energies, from a few to several hundred milliseconds, with a mean typically centred around 100 ms, with energies in the range of $10^{39}-10^{43}$ erg \citep{2017ARA&A..55..261K}. Their spectra are commonly described by cutoff power-law models or by the sum of two blackbody components \citep[e.g][]{2017ApJ...851...17Y,2022ApJS..260...25C}. In some magnetars that went through several periods of bursting activity, the spectral parameters were observed to evolve \citep{2012ApJ...755..150V,2015ApJS..218...11C,2020ApJ...902L..43L}

In addition to emitting short bursts (either isolated or grouped in ``storms" of tens or hundreds in a short period), magnetars can enter long-term outbursts, during which their persistent emission increases by up to several orders of magnitude \citep[e.g.][]{2018MNRAS.474..961C}. These outbursts are often accompanied by active burst periods, with dozens to hundreds of bursts occurring in rapid succession, as observed in sources such as \sgros\ \citep{2006A&A...445..313G,2017ApJ...851...17Y} and \sgrtf\ \citep[e.g.][]{2020ApJ...893..156L,2020ApJ...904L..21Y}.
In contrast, some magnetars have shown little or no bursting activity since their discovery. One example is AX J1818.8--1559, first identified through a single burst detected %in the field of view (FOV) of \integ 
in 2007 \citep{2012A&A...546A..30M}, and with only one possible burst seen  eight years later \citep{2015GCN.18634....1P}.

Several   statistical studies of large samples of magnetar bursts have been published based on data collected with non-imaging instruments on different satellites, including 
International Cometary explorer \citep{1987ApJ...320L.111L,1993ApJ...418..395U,1994A&A...288L..49H,1996Natur.382..518C},
Rossi-XTE \citep{1999ApJ...526L..93G,
2000ApJ...532L.121G,2001ApJ...558..228G,2004ApJ...607..959G,2012ApJ...755....1P},
KONUS instruments on board of Venera 11--14 \citep{1981Ap&SS..75...47M,2001ApJS..137..227A},
KONUS-Wind \citep{1999AstL...25..628M},
CGRO/BATSE \citep{2000ApJ...532L.121G,1994Natur.368..125K}, 
Fermi/GBM \citep{2012ApJ...749..122V,2012ApJ...756...54L,2015ApJS..218...11C}, and
Insight-HXMT \citep{2022ApJS..260...24C,2022ApJS..260...25C}.

Here we report the results of a systematic analysis of magnetar bursts based on data of the INTEGRAL/IBIS instrument, which, thanks to its excellent imaging capabilities in the hard X-ray range, has allowed us to securely identify the magnetar responsible for each detected burst.
We used all the public data available up to October of 2024 to detect bursts from all known magnetars and 
characterise their timing and spectral properties. 

\section{Observations and data reduction}

\integ\ is a satellite of the European Space Agency launched in October 2002, which ceased scientific operations in February 2025. It was designed to perform observations over a broad energy range from 3 keV to 10 MeV. This was achieved thanks to three instruments, the Joint European X-Ray Monitor \citep[JEM-X,][]{2003A&A...411L.231L}, the Imager on Board the INTEGRAL Satellite \citep[IBIS,][]{2003A&A...411L.131U}, and the  SPectrometer on INTEGRAL \citep[SPI,][]{2003A&A...411L..63V} covering the energy ranges 3--35~keV, 15~keV--1 MeV, and 18~keV -- 8 MeV, respectively.    \integ, was able to conduct long and uninterrupted observations, thanks to its highly elliptical orbit with a period of three days until the beginning of 2015 and 64 hours  afterwards \citep{2021NewAR..9301629K}.
Scientific data were acquired during the part of the orbit above the Earth radiation belt \citep[$\sim$40,000 km,][]{2003A&A...411L..43H}, i.e. for about 90\% of time.
All the data were transmitted to Earth in real time, thanks to a continuos telemetry link.

Most of the \integ\  observations were performed using a dithering pattern, in order to optimise imaging with the coded mask instruments and to improve background modelling. Each observation was split in individual segments, referred to as Science Windows (ScWs), typically lasting about $\sim$30 minutes and pointed on a grid of directions centred around the target source. A large part of the observing time was devoted to a survey of the Galactic plane, again using a predefined pattern of ScWs.

We used  data obtained with IBIS, a coded mask telescope with  a wide-field of view (FOV) of 29.1$^{\circ}\times$29.4$^{\circ}$ (at zero sensitivity), including a fully coded, most sensitive central part, of $\sim$ 8$^{\circ}\times$8$^{\circ}$. IBIS is composed of two detectors simultaneously working in different energy ranges:  ISGRI \citep[The \integ\ Soft Gamma-Ray Imager,][]{2003A&A...411L.141L} at lower energies and   PICsIT \citep[][Pixellated Imaging Caesium Iodide Telescope]{2003A&A...411L.149L} at higher energies. 
The time resolution and sensitivity of PICsIT are not adequate to study magnetar bursts. Therefore, we only used  \isgri\ data.

\isgri\ is located at 3.2 meters from the IBIS   coded mask. The mask is composed of gaps and fully opaque elements of tungsten with a thickness of 16 mm. The mask aperture is based on a Modified Uniformly Redundant Array, an optimal pattern that allows resolution of all sources within the field of view. The  images recorded by the ISGRI detector are called  shadowgrams, as they are formed by the superposition of mask shadows cast by the sources in the FoV. The analysis of the shadowgrams, through correlation with the mask pattern, enables the reconstruction of the source positions in the sky.

The \isgri\ detector is composed of 8 modules of $64\times32$ CdTe pixels each, with a pixel size of $4\times4$ mm$^2$ and spaced by 0.6 mm. Including the spacing between the modules, a total shadowgram has a size of $130\times134$ pixels and  a collecting area of 2600 cm$^{2}$. The detector operates nominally in the 15 keV to 1 MeV range, but the effective area is above 900 cm$^2$ in the 20 to 100 keV range and drops below 200 cm$^2$ at E$>200$ keV. The  energy resolution is $\sim$8\% at 100 keV (FWHM). 
\isgri\ works in photon-by-photon mode, providing an excellent time resolution of 61 $\mu$s.% s with an energy resolution of 8\% at 60 keV.

Given  the pitch of the detector pixels (4.6 mm), the size of the coded-mask elements (11.2 mm), and the mask distance (3.2 m), the angular resolution of \isgri\  is  $\sim$12$'$ (FWHM). The  location accuracy depends on the   signal to noise ratio (SNR). It  is typically of $\sim3$ arcmin for sources slightly above the detection threshold, but can reach 1.5 arcmin for SNR above 10 \citep{2013arXiv1302.6915G}.

In the case of bright sources providing a very high count rate, ISGRI data transmission is affected by telemetry saturation. When this occurs, the information on the detected events is not sent to the ground, resulting in gaps in the light curves, typically shorter than a second. For a few bursts affected by this problem, we implemented a correction by linearly interpolating the count rates between saturated periods.   In these cases, the estimated flux and fluence are indicated as   lower limits in the following figures and tables. 

As a result of radiation damage in the CdTe pixels, \isgri\ experienced a gain decrease of $\sim2.6$ \% per year, which was within the expectations \citep{2005ITNS...52.3119L}. In addition, up to 1\%  gain drop was also   caused by strong solar flares (see IBIS Observer’s Manual\footnote{\url{integral.esac.esa.int/AO21/IBIS_ObsMan.pdf}}). As a result, the low threshold of the energy range suitable for spectral analysis evolved with time, reaching a value of 25 keV in 2014, and then almost 40 keV at the end of the mission. Some of the detector pixels can produce spurious events caused by electronic noise, which are flagged and subsequently filtered out in the analysis (only events with  "SELECT\_FLAG" equal to 0 are used).

The source count rates measured by the detector must be corrected for dead time effects. We computed the dead time correction including also the effects induced by the Veto system, by the photons of the on-board calibration source, and  (if enabled) by the tagging of Compton events\footnote{Events producing a coincident signal both in ISGRI and in PICsIt}. These three effects cause a dead time $\epsilon_{hk}$ which can be computed from the housekeeping data of each ScW. The true count rate can then be expressed as:
$N_{true} = N_{rec}\cdot\tau_{tot}$, with $\tau_{tot}$ given by:
\begin{equation}
    \tau_{tot} = \exp \left( \epsilon_{hk} + \frac{N_{rec}}{M} \tau\right)
\end{equation}
\noindent
where $\tau = 0.114$~ms is the fixed dead time during which a pixel cannot register a new event and $M$ is the number of modules illuminated by the source
for at least 20\% of their surface.

The sources in the partially coded field of view do not illuminate the whole detection plane. Therefore, their observed count rate must be corrected for the coding fraction (COD), i.e., the fraction of the detector area exposed to the source. In the following, all reported numbers of burst counts refer to these corrected values unless stated otherwise.

\section{Data analysis}
\subsection{Burst Search}

We considered the 34 sources listed in   Table~\ref{tab:tab1}. Besides the confirmed magnetars, this sample also includes the two rotation-powered pulsars that exhibited magnetar-like emission, PSR J1119-6127 \citep{2016ApJ...829L..21A,2016ApJ...829L..25G} and PSR J1846-0258 \citep{2008Sci...319.1802G,2021ApJ...911L...6B,2024ApJ...976...56S}. For each of the considered sources, we selected all the public ScWs (available at 2024 September 30) pointed within 14.5$^\circ$ from the source position. All the ScWs were screened to remove those affected by   variable background. This was done by fitting a  constant to the light-curve of the whole detector in the full energy range binned at one second, after performing sigma-clipping to exclude possible bursts. ScWs were considered to have good quality and used in the search  for bursts if they had \rchisq$\leq1.3$. This selection resulted in the exclusion of $\sim8\%$ ScWs per source on average.

The Pixel Illumination Factor (PIF) describes the fraction of pixel area  illuminated by a source. Applying a threshold on PIF values allows one to reduce the number of background events and thus to increase the sensitivity of the search.
Before the search, the position of the source was considered to create the PIF matrix for each ScW and filter the list of events with it. In particular, we extracted the source light curves using only pixels with PIF $> 0.5$.

The burst search was then performed on the selected ScWs using the following procedure. The search was done   in eight  time intervals with   logarithmically spaced durations  between 0.01 and 1.28 s. The threshold for each timescale was set at a level above the average (estimated from the sigma-clipped background level in the 1-second light source curve) corresponding to a false positive probability of $10^{-3}$ for each ScW. We used the nominal 15-150 keV energy range in all the ScWs. 

Typically, a single variability event can exceed the thresholds in several timescales and/or bins. After joining adjacent  bins,  only the  trigger with the highest significance was retained.

The triggers found in the light curves were then verified by making images of the corresponding time intervals.  This was done  with an interactive  analysis based on the imaging software developed for the INTEGRAL Burst Alert System \citep[IBAS,][]{2003A&A...411L.291M}. This allowed us to eliminate triggers caused by background variations, sources outside the field of view, or coming from other sources in the field of view but unrelated to the magnetar of interest.

\begin{figure}[t!]
    \centering
    \begin{subfigure}[b]{0.5\textwidth}
         \centering
         \includegraphics[width=\textwidth]{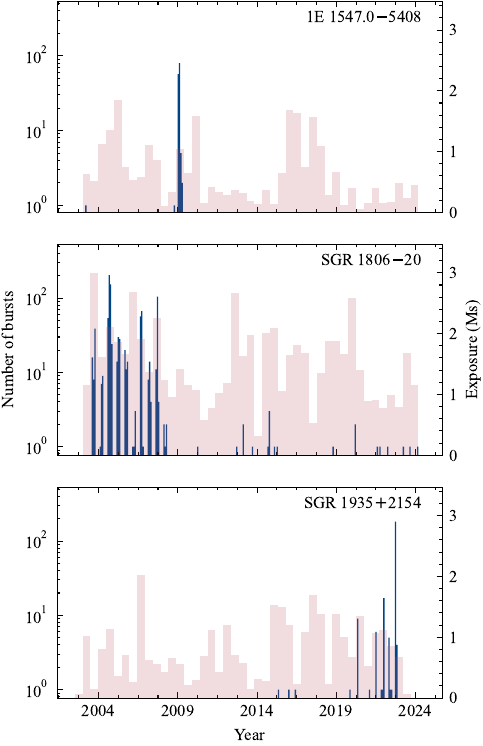}
    \end{subfigure}
\caption{Time distribution of the bursts (blue histograms) and   exposure time (red histograms) of  \onee\ (upper), \sgros\ (middle), and \sgrtf\ (bottom). The bursts are in bins of 30 days, while the exposure is for bins of 180 days.}
\label{fig:fig1}

\end{figure}

In this way, 1349 magnetar bursts were confirmed among more than 75,000 potential triggers found in the light curves search. A   summary of the number of bursts and analysed exposure time for each source    is provided in Table~\ref{tab:tab1}. The largest number of   bursts originated from \sgros, followed by \onee, and \sgrtf. 
The time distributions of the bursts from these three sources are shown in   Fig~\ref{fig:fig1}, where we also indicate the analysed exposure time. 
Additionally, 37 bursts were detected from 18 other magnetars. 

In Table ~\ref{tab:tab2} \footnote{Only the first lines are given here; the whole table is available in the electronic version.} we give the unique  number identifier for each burst, the timing information (see below), the coding fraction (COD), and the fluences in counts (background-subtracted and corrected for the coding fraction).

\begin{deluxetable}{lccrr}
\tablecolumns{5}
\tablecaption{List of the magnetars with coordinates, exposure time and number of detected bursts. \label{tab:tab1}}
\tablehead{\colhead{Source} & \colhead{R.A.} & \colhead{Dec.} & \colhead{Exposure} & \colhead{Bursts} \\
& \multicolumn{2}{c}{(J2000)} & \colhead{(Ms)}}
\startdata
CXO 0100--7211 & 01:00:43 & -72:11:34 & \phn7.96 & 1 \\
4U 0142+614 & 01:46:22 & \phm{-}61:45:03 & 21.35 & 0 \\
SGR 0418+5729 & 04:18:34 & \phm{-}57:32:23 & \phn8.90 & 0 \\
4XMM J045626.3--694723 & 04:56:26 & -69:47:23 & 14.15 & 0 \\
SGR 0501+4516 & 05:01:07 & \phm{-}45:16:34 & \phn9.21 & 8 \\
SGR 0525--66 & 05:26:01 & -66:04:36 & 11.94 & 2 \\
SGR 0755--2933 & 07:55:43 & -29:33:49 & \phn9.55 & 0 \\
1E 1048--5937 & 10:50:07 & -59:53:21 & 23.75 & 2 \\
PSR J1119--6127 & 11:19:14 & -61:27:49 & 25.13 & 1 \\
1E 1547.0--5408 & 15:50:54 & -54:18:24 & 27.09 & 146 \\
Swift J1555.2--5402 & 15:55:09 & -54:03:41 & 27.43 & 0 \\
PSR J1622--4950 & 16:22:45 & -49:50:53 & 28.55 & 0 \\
SGR 1627--41 & 16:35:52 & -47:35:23 & 29.04 & 2 \\
CXOU J164710--455216 & 16:47:10 & -45:52:17 & 31.60 & 3 \\
1RXS J170849--400910 & 17:08:47 & -40:08:52 & 61.82 & 2 \\
CXOU J171405--381031 & 17:14:06 & -38:10:31 & 66.88 & 1 \\
SGR J1745--2900 & 17:45:40 & -29:00:30 & 67.93 & 0 \\
SGR 1801--23 & 18:00:59 & -22:56:49 & 67.04 & 2 \\
SGR 1806--20 & 18:08:39 & -20:24:40 & 65.23 & 934 \\
XTE J1810--197 & 18:09:51 & -19:43:52 & 64.44 & 3 \\
Swift J1818.0--1607 & 18:18:04 & -16:07:32 & 52.89 & 1 \\
AX J1818.8--1559 & 18:18:51 & -15:59:23 & 52.12 & 1 \\
Swift J1822.3--1606 & 18:22:18 & -16:04:27 & 52.20 & 1 \\
SGR J1830--0645 & 18:30:42 & -06:45:17 & 25.28 & 0 \\
SGR 1833--0832 & 18:33:44 & -08:31:08 & 25.50 & 3 \\
Swift J1834.9--0846 & 18:34:52 & -08:45:56 & 25.62 & 0 \\
1E 1841--045 & 18:41:19 & -04:56:11 & 28.21 & 1 \\
AX J1845.0--0258 & 18:44:55 & -02:56:53 & 30.51 & 0 \\
PSR J1846--0258 & 18:46:25 & -02:58:30 & 30.57 & 2 \\
3XMM J185246.6+003317 & 18:52:47 & \phm{-}00:33:18 & 31.18 & 0 \\
SGR 1900+14 & 19:07:14 & \phm{-}09:19:20 & 27.57 & 1 \\
SGR 1935+2154 & 19:34:56 & \phm{-}21:53:48 & 32.90 & 232 \\
SGR 2013+34 & 20:13:57 & \phm{-}34:19:48 & 27.98 & 0 \\
1E 2259+586 & 23:01:08 & \phm{-}58:52:45 & 25.24 & 0 \\
\enddata
% \tablenotetext{a}{J2000}
% \tablenotetext{b}{Ms}

\end{deluxetable}

\begin{figure}[t!]
    \centering
    \begin{subfigure}[b]{0.32\textwidth}
         \centering
         \includegraphics[width=\textwidth]         {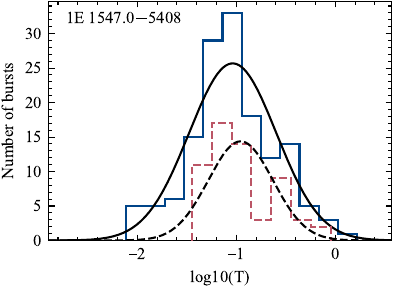}
    \end{subfigure}
    \hfill
    \begin{subfigure}[b]{0.32\textwidth}
         \centering
         \includegraphics[width=\textwidth]{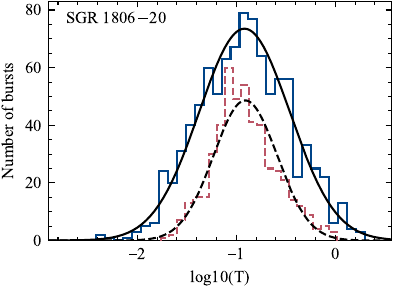}
    \end{subfigure}
    \hfill
        \begin{subfigure}[b]{0.32\textwidth}
         \centering
         \includegraphics[width=\textwidth]{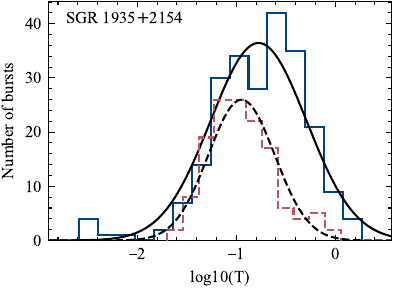}
    \end{subfigure}

    \begin{subfigure}[b]{0.32\textwidth}
         \centering
         \includegraphics[width=\textwidth]{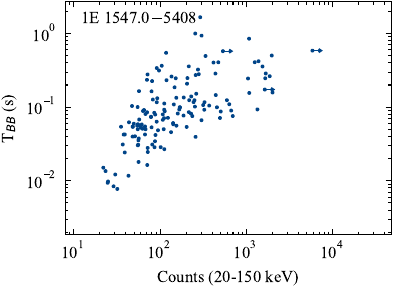}
    \end{subfigure}
    \hfill
    \begin{subfigure}[b]{0.32\textwidth}
         \centering
         \includegraphics[width=\textwidth]{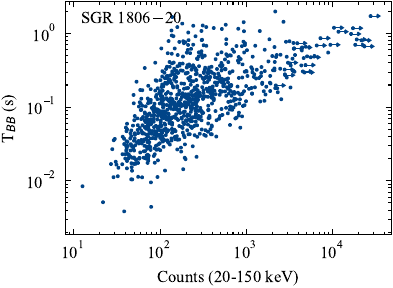}
    \end{subfigure}
    \hfill
    \begin{subfigure}[b]{0.32\textwidth}
         \centering
         \includegraphics[width=\textwidth]{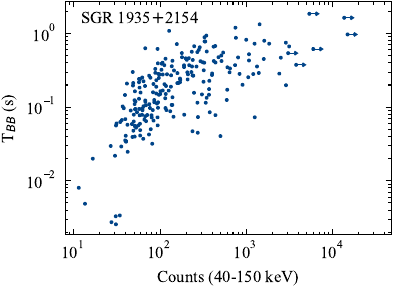}
    \end{subfigure}
\caption{Top panels: Distributions of the burst durations fitted with lognormal functions for \protect\onee\  (left), \protect\sgros\ (middle), and \protect\sgrtf\ (right). Solid lines refer to \Tbb\ and dashed lines to \Tn. Bottom panels:  \Tbb\ duration as a function of the  number of counts for the same three magnetars. }
\label{fig:fig2}

\end{figure}

\subsection{Timing Analysis}
\label{sec:TA}
The timing characterisation of the bursts was performed using the Bayesian Blocks (BB) method \citep{1998ApJ...504..405S,2013ApJ...764..167S}, in particular to estimate their  duration and to define the time intervals for the spectral analysis.
This is a non-parametric technique that characterises intensity variations from an unbinned  list of event arrival times by dividing the data into blocks, each of which shows no statistically significant variations in count rate. This ensures that each block is   consistent with  a period of  stable intensity. Only statistically significant changing points are introduced, based on a parameter, p$_0$, that represents the false-positive probability threshold used to determine the change points in the data. The BB analysis was done on a 30 s long time interval centered on each burst and we used p$_0$=0.05.

To allow comparison with results in the literature, we also computed the bursts \Tn, which 
is the duration of the time interval containing 90\% of the burst fluence (setting T$_{start}$ and T$_{stop}$ at times corresponding to 5\% and 95\% percent of the counts, respectively). 
The background level for the \Tn\  computation was derived from the BB results. 
In particular, we  selected the two blocks with the longest duration, one before and one after the burst, and took their average as the background count rate. 
The background-subtracted integral distribution of burst counts was then fitted with an error function plus a linear term to take into account residual background fluctuations that were present in some cases.
However, this method had limitations, particularly when dealing with very faint bursts or certain burst shapes, which   prevented a correct fit, usually causing unnaturally large \Tn. Thus, each burst light curve was visually inspected and rejected in case of incorrect estimation. \Tn\ could be estimated for 692 bursts. No alternative \Tn\ estimation methods were applied to the problematic cases, to maintain consistency throughout the analysis.

\subsection{Spectral analysis}

We performed a spectral analysis for the bursts with at least 100 observed counts in the 20--150 keV range for bursts before January 2018, or in the  40--150 keV range after this date. 
For the spectral analysis, we used the pyXspec \citep{2021ascl.soft01014G} version 2.1.4 with the Xspec version 12.14.1 \citep{1996ASPC..101...17A}.
Burst spectra were extracted using the standard pipeline with OSA 11.2 \citep{2003A&A...411L.223G} using custom-defined Good Time Intervals (GTI) that were based on the \Tbb\ durations of the bursts. If a burst was affected by telemetry saturation, the saturated periods were excluded from  the GTIs.
The energy range E$_{min}$--E$_{max}$  used for the  spectral analysis depended 
on the date of the burst, in order  to take into account the evolution of the lower energy threshold. We used  E$_{min}$ values  between   $20$ and $32$ keV, depending on the observation date,  and E$_{max} =  300$ keV. The energy range was divided in  16 or in 13 logarithmic channels. 
The spectra were rebinned to obtain a minimum of 20 counts in each bin, and 5\% systematic errors were added to the data. 
We considered two models, a simple power law and a power law with exponential cut-off (often called Comptonized model in the literature), defined as follows: 
\begin{itemize}
    \item  power law (PL): $F(E) = k E^{\alpha}$  ph cm$^{-2}$ s$^{-1}$ keV$^{-1}$,
    \item cut-off power law (CPL):  $F(E) = k E^{\alpha} {\rm exp}(-(E(2+\alpha)/E_{peak}))$  ph cm$^{-2}$ s$^{-1}$ keV$^{-1}$.
\end{itemize}

For the bursts with the highest number of counts we performed also a time-resolved analysis  by extracting spectra from different time intervals
(only   bursts from \sgros\ and \sgrtf\ satisfied this requirement). The analysis was done as described above,  using the  20--300 keV and 32--300 keV energy ranges, for \sgros\ and \sgrtf, respectively, divided in  13 bins.

The selection of the burst segments was based on the Bayesian Blocks; however, in some cases, the segmentation was done manually only based on the visual inspection of the light curve and Bayesian block structure. If the block contained sufficient counts, it was divided into segments of approximately equal duration; conversely, adjacent blocks were merged when individual segments lacked sufficient statistics.

\section{Results}

\subsection{Burst durations}

The derived durations of all the bursts are given in Table~\ref{tab:tab2} and the    distributions of \Tbb\  and \Tn\ for \sgros, \onee, and \sgrtf\   are shown in the top panels of  Fig.~\ref{fig:fig2}. 
The bottom panels of the same figure show  the duration \Tbb\ as a function of the number of counts.
The distributions of both \Tbb\  and \Tn\ are well fit with   log-normal curves\footnote{Two particularly long bursts emitted  by \sgros\ on 2004 October 5 (n.387 and n.414) have been excluded from this analysis, because they were affected by telemetry saturation and most likely involved the superposition of several bursts.  
They have been  analysed  in detail by \citet{2006A&A...445..313G}.}, 
as also confirmed using the Kolmogorov-Smirnov test.
The burst average durations  and the best fit parameters of the log-normal fits (mean $\mu$ and standard deviation $\sigma$) for the three sources are given in Table~\ref{tab:tab3}.

\begin{figure}
    \centering
    \includegraphics[width=0.5\linewidth]{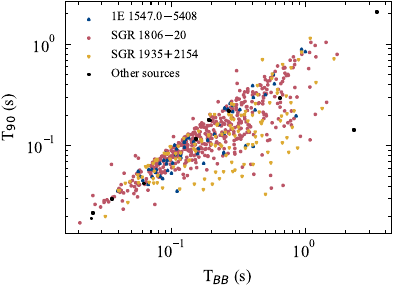}
    \caption{\protect\Tn\  versus \protect\Tbb\   of all the bursts for which    \protect\Tn\ could be estimated.}
    \label{fig:fig3}
\end{figure}

As it is shown in Fig.~\ref{fig:fig3}, \Tbb\  is generally longer than \Tn. 
Indeed,  \Tbb\ is an estimate of the whole burst duration and the BB algorithm is more sensitive to measure faint emission extending before and/or after the main burst peaks.
The  durations estimated with \Tbb\ suggest a tendency for shorter bursts in \onee\ and longer ones in \sgrtf . On the other hand the  parameters of the \Tn\ distributions are  very similar for the three sources. This apparent discrepancy is probably due to the limitations in deriving reliable  \Tn\ described in section~\ref{sec:TA}.  In fact \Tn\ could not be  measured for the faintest bursts, therefore a direct comparison between the average \Tn\ and \Tbb\  is not straightforward because they refer to different samples.

\begin{deluxetable}{rlllrrrrr}
\tablecolumns{10}
\tablecaption{Detected bursts from magnetars.\label{tab:tab2}}
\tablehead{\colhead{No} & \colhead{Source} & \colhead{T$_{start}$} & \colhead{\Tbb} & \colhead{\Tn} & \colhead{COD} & \colhead{CTS$_{20-40}$}  & \colhead{CTS$_{40-100}$} & \colhead{CTS$_{100-150}$}\\ &  & \colhead{(UTC)} &  \colhead{(ms)} & \colhead{(ms)}}
\startdata
0 & 1E 1547.0-5408 & 2003-03-14 00:09:42.553 & 68.0 & \nodata & 0.74 & 1.7 & 39.0 & 16.5 \\
1 & SGR 1627-41 & 2003-03-25 02:45:59.024 & 111.7 & \nodata & 1.00 & 0.0 & 27.0 & 19.7 \\
2 & 1RXS J170849-400910 & 2003-03-26 05:48:55.080 & 121.9 & \nodata & 0.98 & 4.3 & 33.3 & 18.3 \\
3 & SGR 1833-0832 & 2003-04-26 11:34:19.751 & 532.9 & \nodata & 0.67 & 26.8 & 50.6 & 9.1 \\
4 & 1E 1048-5937 & 2003-07-04 01:53:56.652 & 2303.3 & 142.3 & 1.00 & 33.5 & 70.8 & 18.2 \\
5 & SGR 1806-20 & 2003-08-23 17:32:11.718 & 216.1 & 121.8 & 0.10 & 104.0 & 190.8 & 29.8 \\
6 & SGR 1806-20 & 2003-08-23 22:05:01.450 & 129.5 & 91.3 & 0.40 & 171.2 & 138.1 & 4.8 \\
7 & SGR 1806-20 & 2003-08-23 22:05:01.776 & 87.6 & \nodata & 0.40 & 85.5 & 97.6 & 10.1 \\
8 & SGR 1806-20 & 2003-08-24 15:01:11.846 & 213.0 & 189.9 & 0.86 & 48.4 & 55.7 & 11.8 \\
9 & SGR 1806-20 & 2003-08-24 15:30:44.152 & 254.8 & 177.2 & 0.86 & 240.0 & 172.6 & 19.8 \\
\enddata
\tablecomments{The whole table is available online  in  machine-readable format.
A portion is shown here for guidance regarding its form and content.}
\end{deluxetable}
\begin{deluxetable}{lllll}
\tablecolumns{5}
\tablecaption{Properties of the burst durations.\label{tab:tab3}}
\tablehead{\colhead{Source} &   & \colhead{Average (ms) } & \colhead{10$^{\mu}$ (ms)} & \colhead{$\sigma$ }}
\startdata
\onee & \Tbb & $157.3$ & $92.5\pm7.8$ & $0.44\pm0.03$ \\
 & \Tn & $154.4$ & $110.2\pm10.6$ & $0.32\pm0.03$ \\
\sgros  & \Tbb & $206.6$ & $120.7\pm4.2$ & $0.46\pm0.01$ \\
  & \Tn & $165.9$ & $123.4\pm4.2$ & $0.32\pm0.01$ \\
\sgrtf & \Tbb & $274.0$ & $168.5\pm12.3$ & $0.48\pm0.02$ \\
 & \Tn & $159.1$ & $113.1\pm7.5$ & $0.33\pm0.02$ \\
\enddata
\tablecomments{$\mu$ and $\sigma$ are the mean and standard deviation of lognormal fits to the distributions of durations.}

\end{deluxetable}

\subsection{Burst spectra}
\begin{deluxetable}{llccc|cccc}
\tablecolumns{9}
\tablecaption{Results of the timed-averaged spectral analysis of the bursts. \label{tab:tab4}}
\tablehead{  & & \multicolumn{3}{c}{Power law} & \multicolumn{4}{c}{Cutoff power law}\\ \colhead{No} & \colhead{Source}  &  \colhead{$\alpha$} & \colhead{\rchisq(dof)} & \colhead{Flux} & \colhead{$\alpha$} & \colhead{\Ep}& \colhead{\rchisq(dof)}  & \colhead{Flux} \\ \multicolumn{4}{c}{} & \colhead{($10^{-7}$ \flux)} & &\colhead{(keV)} & & \colhead{($10^{-7}$ \flux)}}
\startdata
6 & SGR 1806-20 & $-2.2 \pm 0.2$ & 1.6(8) & $2.9 \pm 0.4$ & $-0.5^{+1.3}_{-1.0}$ & $46.3^{+8.6}_{-9.6}$ & 1.3(7) & $3.0^{+0.4}_{-0.5}$\\
9 & SGR 1806-20 & \nodata & \nodata & \nodata & $-0.4^{+1.0}_{-0.9}$ & $40.0^{+5.2}_{-8.8}$ & 1.5(6) & $1.6 \pm 0.2$\\
14 & SGR 1806-20 & $-2.1 \pm 0.3$ & 0.8(6) & $3.3 \pm 0.6$ & \nodata & \nodata & \nodata & \nodata \\
17 & SGR 1806-20 & $-2.1 \pm 0.2$ & 0.8(9) & $1.6 \pm 0.2$ & \nodata & \nodata & \nodata & \nodata \\
19 & SGR 1806-20 & $-2.1 \pm 0.1$ & 0.8(9) & $2.1 \pm 0.2$ & \nodata & \nodata & \nodata & \nodata \\
28 & SGR 1806-20 & $-2.6 \pm 0.1$ & 1.5(8) & $4.4 \pm 0.4$ & \nodata & \nodata & \nodata & \nodata \\
32 & SGR 1806-20 & \nodata & \nodata & \nodata & $-0.6^{+0.5}_{-0.9}$ & $24.9^{+3.2}_{-10.8}$ & 0.6(8) & $12.0 \pm 0.8$\\
34 & SGR 1806-20 & $-2.8 \pm 0.2$ & 1.4(7) & $2.2 \pm 0.3$ & \nodata & \nodata & \nodata & \nodata \\
35 & SGR 1806-20 & $-2.4 \pm 0.2$ & 1.3(7) & $1.1 \pm 0.2$ & \nodata & \nodata & \nodata & \nodata \\
36 & SGR 1806-20 & $-2.1^{+0.3}_{-0.4}$ & 0.7(4) & $1.3 \pm 0.3$ & \nodata & \nodata & \nodata & \nodata \\
\enddata
\tablecomments{The whole table is available online  in  machine-readable format.
A portion is shown here for guidance regarding its form and content.}

\end{deluxetable}

In total, 535 bursts had enough counts for spectral analysis. 
The   results of the spectral fits for each burst are listed in Table~\ref{tab:tab4}. The power law model provided an acceptable fit for 313 bursts, 
while a CPL was required in 114 cases. The other bursts  could not be well fitted with either model
(due to saturation problems or because the fit parameters were completely unconstrained).
For 54 of the bursts adequately fit by a PL, the addition of a cut-off provided a statistically significant improvement,  as determined by an F-test with a p-value below 0.15. 
Thus, in summary, we have a sample of 168 bursts for which $\alpha$ and \Ep\ could be derived, although for some of them (30), the peak energy E$_{peak}$ 
was poorly constrained (the lower bound of its confidence interval was below 10 keV).

\begin{figure}[ht!]
    \centering
    \begin{subfigure}[b]{0.32\textwidth}
         \centering
         \includegraphics[width=\textwidth]{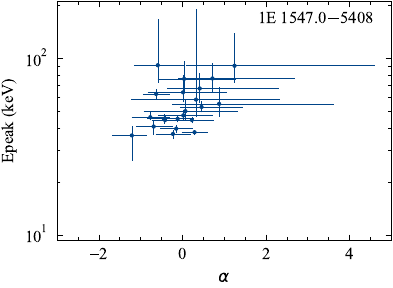}
    \end{subfigure}
    \hfill
    \begin{subfigure}[b]{0.32\textwidth}
         \centering
         \includegraphics[width=\textwidth]{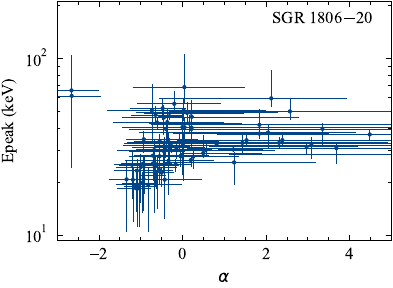}
    \end{subfigure}
    \hfill
    \begin{subfigure}[b]{0.32\textwidth}
         \centering
         \includegraphics[width=\textwidth]{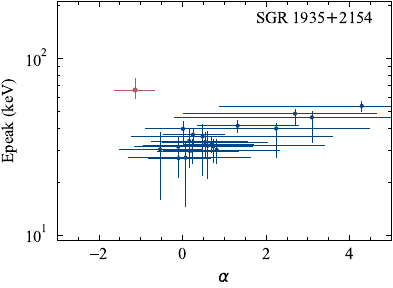}
    \end{subfigure}

    \begin{subfigure}[b]{0.32\textwidth}
         \centering
         \includegraphics[width=\textwidth]{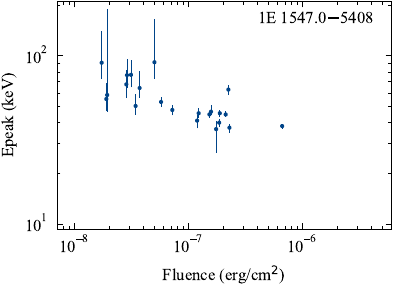}
    \end{subfigure}
    \hfill
    \begin{subfigure}[b]{0.32\textwidth}
         \centering
         \includegraphics[width=\textwidth]{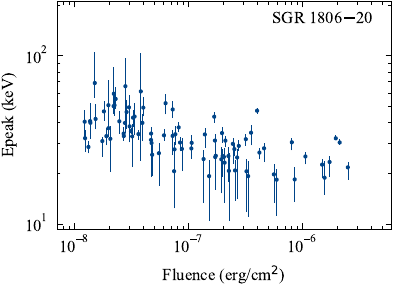}
    \end{subfigure}
    \hfill
    \begin{subfigure}[b]{0.32\textwidth}
         \centering
         \includegraphics[width=\textwidth]{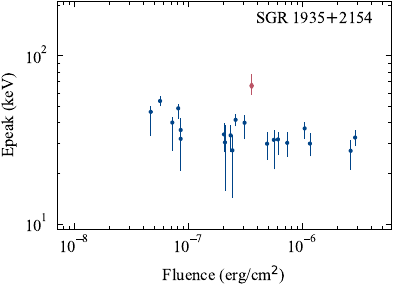}
    \end{subfigure}
    \caption{E$_{peak}$ versus $\alpha$ (upper panels) and versus fluence (bottom panels) for the spectra fitted by the cut-off power law model.  The figure show results for  three sources: \onee\ (left), \sgros\ (middle), and \sgrtf\ (right). The fluence refers to the 30--150 keV range. The  2020 April 28 burst  from \sgrtf\ with associated FRB-like emission is indicated in red. }
    \label{fig:fig4}
\end{figure}

We computed the error-weighted average values of the best fit spectral parameters.  The average values of the CPL model were then used to estimate Energy Conversion Factors (ECFs) from the count rates ($20-150$ keV and $40-150$ keV)  to the $30-150$ keV flux in physical units.
The average spectral parameters and the ECFs  are listed in Table~\ref{tab:tab5}.
The relation between $\alpha$ and E$_{peak}$ is plotted in the top panels of Fig.~\ref{fig:fig4}, while the bottom panels  of the same figure show how the spectral parameters depend on the burst fluence. 

These figures show evidence for some correlations between the parameters. To quantify this taking into account the measurement errors, we computed the Spearman's rank correlation coefficient $\rho$ with a Monte Carlo approach that combines bootstrap resampling with perturbations \citep{2014arXiv1411.3816C} drawn from the split-normal error distributions for each data point. 
We found a very significant anti-correlation between \Ep\ and fluence for \sgros, with a mean $\rho=-0.48\pm0.07$ and a mean $z$-score of $7.1\pm1.0$ from 100,000 Monte Carlo tests.
For this source, there is also marginal evidence for a correlation between $\alpha$ and \Ep\ ($\rho=- 0.31\pm0.09$,   $z$-score=$3.4\pm1.1$).
These correlations are less significant for the other two magnetars (see the $\rho$ and  $z$-score values in Table~\ref{tab:tab6}).

\begin{deluxetable}{lrrrrr}
\tablecolumns{6}
\tablecaption{Average spectral parameters and  Energy Conversion Factors for the cut off power law model.
\label{tab:tab5}}
\tablehead{\colhead{Source} & \colhead{$\alpha_{PO}$} & \colhead{$\alpha_{CPL}$} & \colhead{\Ep\tablenotemark{a}} & \colhead{ECF\tablenotemark{b}$_{20-150}$} & \colhead{ECF\tablenotemark{c}$_{40-150}$}}
\startdata
\onee &  $-1.78\pm0.03$ & $-0.32\pm0.11$ & $43.3\pm0.7$ & 1.1 & 2.2 \\
\sgros& $-2.42\pm0.01$ & $-0.60\pm0.07$ & $32.4\pm0.4$ & 0.9 & 2.2 \\
\sgrtf & $-2.68\pm0.03$ & $0.03\pm0.25$  & $37.7\pm1.4$ & \nodata  & 1.9 \\
\enddata
\tablenotetext{a}{keV}
\tablenotetext{b}{from counts in the $20-150$ keV range to $10^{-10}$ \fluence\ in the  $30-150$ keV range.}
\tablenotetext{c}{from counts in the $40-150$ keV range to $10^{-10}$ \fluence\ in the  $30-150$ keV range.}
\end{deluxetable}

We performed time-resolved spectral analysis for 21 bursts (15   from \sgrtf\ and 6 from \sgros).
First we fitted the spectra with the  power law model and  no significant spectral evolution was observed in the photon index.
Therefore, we then used the cutoff power law model, keeping  $\alpha$ tied across all the time intervals of each burst, while \Ep\ and the normalization were allowed to vary independently.

\begin{figure}[t!]
    \centering
         \begin{subfigure}[b]{0.49\textwidth}
          \centering
          \includegraphics[width=\textwidth]{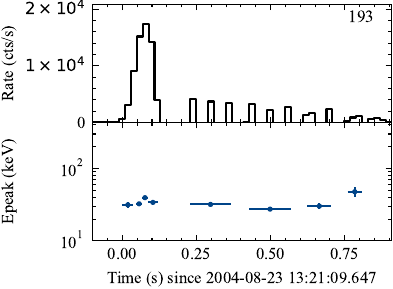}
     \end{subfigure}
      \hfill
     \begin{subfigure}[b]{0.49\textwidth}
           \centering
           \includegraphics[width=\textwidth]{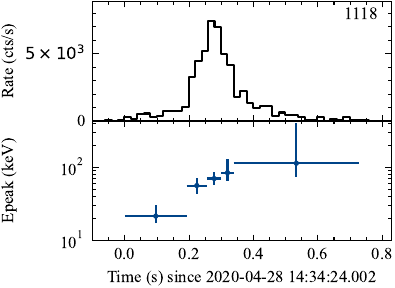}
     \end{subfigure}

     \begin{subfigure}[b]{0.49\textwidth}
          \centering
          \includegraphics[width=\textwidth]{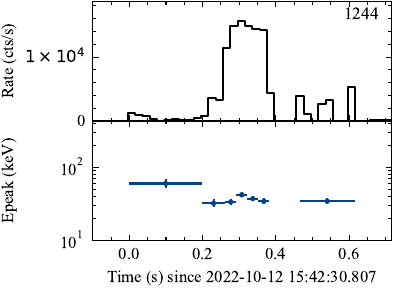}
     \end{subfigure}
     \hfill
     \begin{subfigure}[b]{0.49\textwidth}
          \centering
          \includegraphics[width=\textwidth]{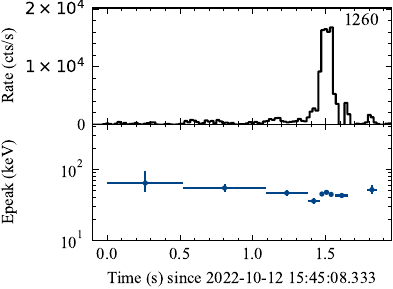}
     \end{subfigure}

     \begin{subfigure}[b]{0.49\textwidth}
          \centering
          \includegraphics[width=\textwidth]{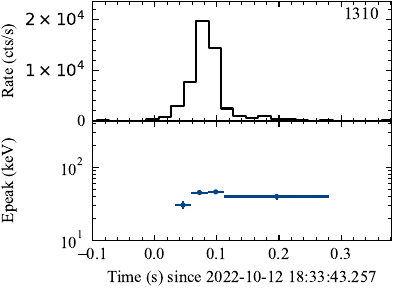}
     \end{subfigure}
     \caption{Time resolved spectral analysis of one burst (n. 193) from \sgros\ and four bursts from \protect\sgrtf. The upper panel in each  figure shows the background subtracted light curve of the burst, while the bottom panel shows the time evolution of $E_{peak}$.
     The gaps in the light curve of burst n. 193 are due to telemetry saturation. 
     \label{fig:fig5}}
\end{figure}

In the majority of cases, \Ep\ was found to be consistent,  within the errors, with a constant value. Only in five cases, we detected a marginally significant variation (p-value $< 0.05$, for bursts no 193, 1118, 1244, and 1310  and  p-value $< 0.1$ for no 1260). These are shown in Fig.~\ref{fig:fig5} and their spectral parameters are given in Table \ref{tab:tab7}.

\subsection{Log N -- Log S}

\begin{figure}[ht]
    \centering
    \begin{subfigure}[b]{0.49\textwidth}
         \centering
         \includegraphics[width=\textwidth]{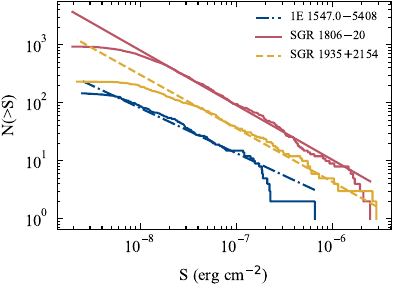}
    \end{subfigure}
    \caption{ LogN-LogS distributions of bursts of \onee\ (grey), \sgros\ (yellow), and \sgrtf\ (blue). The best fit power laws are indicated by the dashed lines. The fluence S refers to the 30--150 keV energy range.\label{fig:fig6}
}
\end{figure}

For each of the three magnetars with a large number of detected bursts we calculated the integral distribution of burst fluences. The fluences, S, in the 30--150 keV energy range were computed from the  number of counts using the ECF values reported in Tab.~\ref{tab:tab5}. 
These distributions (LogN-LogS) are plotted in Fig.~\ref{fig:fig6}. 
In order to compare these distributions with other published results, we fitted them with a power law model ($N(>S)\propto S^{-\beta}$). To avoid potential biases resulting from least-square fitting of the binned differential distributions, we followed a maximum-likelihood approach to derive from the unbinned data the power law slope $\beta$  and the minimum fluence $S_{min}$ above which a power-law is a good description of the data \citep{1970ApJ...162..405C,2009SIAMR..51..661C}.  The results obtained for the three sources are given  in Table~\ref{tab:tab6}.
The deviation from a power-law trend visible at low fluences is due to the decreasing efficiency in the detection of the faintest bursts. In fact the highest $S_{min}$ value was obtained  for \sgrtf ,  that was active during the latest part of the mission, when the \isgri\ sensitivity decreased.

\begin{deluxetable}{lllcccc}
\tablecolumns{7}
\tablecaption{Spearman's rank correlations and best fit  parameters of the fluence distributions ($N(>S)\propto S^{-\beta}$, for $S>S_{min}$) \label{tab:tab6}}
    \tablehead{ \colhead{Source} & \colhead{$\rho(E_p,S)$} & \colhead{$z$--score} & \colhead{$\rho(E_p,\alpha)$} & \colhead{$z$--score} & \colhead{$\beta$} & \colhead{$S_{min}$} \\ &  & & & & & \colhead{($10^{-8}$ erg cm$^{-2}$)}}
\startdata
\onee & $-0.51 \pm 0.19$ & $2.5\pm1.2$ & $0.22 \pm 0.19$ & $1.0\pm1.1$ &  $0.76\pm0.04$ & $0.53$\\
\sgros & $-0.48\pm0.07$ & $7.1\pm1.0$ & $0.31 \pm 0.09$ & $3.4\pm1.1$ &  $0.95\pm0.06$ & $1.89$ \\
\sgrtf & $-0.39 \pm 0.15$ & $2.2\pm0.9$ & $0.38 \pm 0.17$ & $2.1\pm1.1$ &  $0.92\pm0.10$ & $3.95$ \\
\sgrtf\tablenotemark{a} & $-0.45 \pm 0.15$ & $2.5\pm1.0$ & $0.46 \pm 0.16$ & $2.6\pm1.1$ &  \nodata & \nodata \\
\enddata
\tablenotetext{a}{Excluding burst coincident with FRB 200428}
\end{deluxetable}

\section{Discussion}
\subsection{ \onee }
The active period of \onee\ spanned from October 2008 to April 2009 and included three distinct bursting episodes \citep[e.g.][]{2011ApJ...739...94S,2012ApJ...749..122V}, in which \integ\ detected 1, 106, and 38 bursts, respectively. \citet{2010A&A...510A..77S}  studied the January 2009 activity using both the Anti-Coincidence Shield (ACS) of the spectrometer and ISGRI, reporting over 200 bursts with ACS and 84 with ISGRI. They measured a burst duration distribution centred at $68 \pm 3$ ms using ACS data. In comparison, our ISGRI analysis during the same period yielded a longer duration distribution.
Although discovered several years earlier, \onee\ was suggested as a magnetar candidate only in 2007, based on its association with SNR G327.24-–0.13 and the discovery of radio pulsations \citep{2007ApJ...667.1111G,2007ApJ...666L..93C}. The first bursts from this source were reported by Swift during the 2008 October outburst \citep{2011ApJ...739...94S}, but a burst detected by  INTEGRAL on 14 March 2003 (no 0 in Table~\ref{tab:tab2}) indicates that \onee\ was also active at earlier times.
\citet{2012ApJ...755..150V} reported spectral evolution between the first and third active periods. In our case, such evolution, between the second and the third period, could not be assessed due to the limited number of bursts with sufficient counts for detailed spectral fitting during the third period, as only two bursts could be fitted with the cut-off power-law model. 
The fluence distribution of \onee\ bursts was previously derived using Swift/XRT data in the 1-10 keV range by \citet{2011ApJ...739...94S}, who found a power law  index  $\beta=0.6\pm0.1$.
A similar slope was obtained at higher energies (8-200 keV, $\beta=0.7\pm0.2$) with the Fermi/GBM instrument \citep{2012ApJ...749..122V} and with the INTEGRAL SPI/ACS instrument at E$>$80 keV ($\beta=0.75\pm0.06$, \citet{2009ApJ...696L..74M}).
All these results are in agreement we the    value $\beta=0.76\pm0.05$ derived in our analysis.

The \onee\ bursts detected by IBIS   are on average harder  than those seen from the other two magnetars (see Fig.~\ref{fig:fig4} and Table \ref{tab:tab5}). We obtained an average value of \Ep = 43.7$\pm$0.7 keV  consistent with that  obtained  with Fermi/GBM during the January 2009 activity period \citep{2012ApJ...756...54L}, which is the one that dominates our sample (see Fig.~\ref{fig:fig1}).

\subsection{\sgros }
\sgros\ was the most active source during the \integ\ mission, with a total of 934 bursts and, providing the largest sample in our data, it dominates the overall burst statistics. The bursting activity of \sgros\ began with a high-bursting period over a year before the Giant Flare in 2004 \citep{2005ApJ...628..938M}.  After April 2008 only 20 bursts were   detected in our data, and also other satellites detected only sporadic bursts \citep{2015ApJS..218...11C,2017ApJ...851...17Y}.
The fluence distribution of the bursts detected from \sgros\  follows a power law of slope 0.95$\pm$0.06 for fluences above $\sim2\times10^{-8}$ erg cm$^{-2}$. This agrees well with the previous value, $\beta=0.91\pm0.09, $ obtained with a smaller subset of these bursts \citep{2006A&A...445..313G}. Flatter slopes ($\beta\sim0.5$) were found for the bursts detected with the $RXTE$ satellite \citep{2012ApJ...755....1P,2000ApJ...532L.121G}, which operated at lower energies (2-60 keV).

\subsection{\sgrtf }
All the bursts detected for \sgrtf\ occurred between 2015 and 2022.  This sample includes the burst with associated FRB-like radio emission on 2020 April 28 \citep{2020ApJ...898L..29M,2020Natur.587...54C,2020Natur.587...59B}, but the period of high bursting activity that preceded this event \citep{2020ApJ...904L..21Y,2021ApJ...916L...7K} was not observed by \integ.  
Most of the bursts were detected during the outburst of October 2022 (186 bursts between October 10 and 16). 
This outburst was also covered by the Fermi satellite, which reported 113 bursts \citep{2025ApJS..276...60R}.
The bursts detected by the Fermi/GBM instrument were on average shorter  (log-normal distribution centred at $\sim120$ ms) and had lower $E_{\rm peak}$ values ($\sim26$ keV) than those observed by \integ , %($168.51 \pm 5.34$ ms and $38.1\pm1.4$ keV), 
but note  that the two studies were based on different samples of bursts that only partially overlapped.
These authors found a fluence distribution with $\beta=1.2\pm0.2$ for the October 2022 bursts, steeper than what observed with Fermi/GBM in the other activity periods of \sgrtf , and consistent within the errors with the value we found ($\beta=0.92\pm0.10$).

Our spectral results for   the April 2020 burst with associated FRB-like radio emission (burst n. 1118 in our sample, see Fig.~\ref{fig:fig5}) agree with those reported in the previous analysis of these data \citep{2020ApJ...898L..29M},
when one accounts for the different energy ranges of the reported fluxes and the slightly different integration times. In particular, we confirm that the  time averaged peak energy was greater than that of bursts of similar fluence (Fig.~\ref{fig:fig4}),  and that the spectrum hardened with  time with $E_{peak}$ evolving from $\sim$20 to $\sim100$ keV (Fig.~\ref{fig:fig5}).   
The peculiar spectral properties of this bursts were seen also in the data obtained with the Insight-HXMT satellite over the broad energy range of 1--250 keV \citep{2021NatAs...5..378L} and with Konus-Wind \citep{2021NatAs...5..372R}. Several authors speculated that this is related to the unique radio burst emitted during this event \citep{2020Natur.587...59B, 2020Natur.587...54C}, although a clear interpretation of such a connection is still a matter of debate 
\citep[see, e.g.,][]{2020ApJ...904L..15I,2021NatAs...5..378L,2025ApJ...988..274W}.

Radio bursts from \sgrtf\ were detected also on 2020 April 30 \citep{2020ATel13699....1Z}, on 2020 September 2 \citep{2020ATel14186....1A},
on 2020 October 8 \citep{2020ATel14074....1G,2023arXiv231016932G},
on 2022 October 14 and 21 \citep{2022ATel15697....1M,2022ATel15681....1D},
and on 2022 December 1 \citep{2023arXiv231016932G},
but unfortunately they were not  covered by \integ\ data.

\startlongtable
\begin{deluxetable}{cccrrr}
\tablecolumns{6}
\tablecaption{Parameters of time-resolved spectral analysis of the selected bursts. Segments were simultaneously fitted with tied alpha.\label{tab:tab7}}
\tablehead{\colhead{No} & \colhead{T$_{start}^{a}$} & \colhead{T$_{stop}^{a}$} & \colhead{$\alpha$} & \colhead{\Ep} & \colhead{Flux} \\ &  & & & \colhead{(keV)} & \colhead{($10^{-7}$\flux)}}
\startdata
193 & 0.000 & 0.808 & $0.8^{+0.6}_{-0.3}$ & $32.4^{+1.5}_{-0.6}$  & $16.4 \pm 0.6$ \\
 & 0.000 & 0.037 & $2.3^{+0.3}_{-0.9}$ & $31.5^{+2.7}_{-3.0}$ & $7.4^{+1.1}_{-0.9}$ \\
 & 0.047 & 0.066 & \nodata & $32.6^{+1.3}_{-1.8}$ & $29.4^{+2.3}_{-2.2}$ \\
 & 0.066 & 0.086 & \nodata & $39.6^{+1.4}_{-1.6}$ & $37.9^{+2.6}_{-2.4}$ \\
 & 0.086 & 0.120 & \nodata & $34.0^{+1.2}_{-1.5}$ & $33.2^{+2.2}_{-2.1}$ \\
 & 0.229 & 0.365 & \nodata & $31.9^{+1.8}_{-2.1}$ & $19.8^{+2.0}_{-1.8}$ \\
 & 0.429 & 0.568 & \nodata & $27.5^{+1.5}_{-2.0}$ & $12.7^{+1.4}_{-1.3}$ \\
 & 0.624 & 0.704 & \nodata & $30.3^{+2.3}_{-2.9}$ & $9.0^{+1.2}_{-1.1}$ \\
 & 0.760 & 0.808 & \nodata & $47.8^{+6.7}_{-7.6}$ & $1.6^{+0.4}_{-0.3}$ \\
1118 & 0.000 & 0.728 & $-1.1 \pm 0.5$ & $66.4^{+11.4}_{-7.7}$  & $6.0 \pm 0.3$ \\
 & 0.000 & 0.192 & $-1.1^{+0.4}_{-0.6}$ & $21.6^{+4.3}_{-9.3}$ & $1.5^{+0.3}_{-0.2}$ \\
 & 0.192 & 0.255 & \nodata & $56.5^{+12.6}_{-15.9}$ & $13.3^{+1.3}_{-1.2}$ \\
 & 0.255 & 0.300 & \nodata & $70.5^{+12.9}_{-17.2}$ & $29.5^{+2.2}_{-2.0}$ \\
 & 0.300 & 0.338 & \nodata & $84.7^{+19.1}_{-45.6}$ & $20.2^{+1.8}_{-1.7}$ \\
 & 0.338 & 0.728 & \nodata & $114.0^{+40.0}_{-297.7}$ & $2.5^{+0.3}_{-0.2}$ \\
1244 & 0.000 & 0.616 & $0.2^{+1.1}_{-0.7}$ & $30.1 \pm 4.9$  & $12.7 \pm 0.4$ \\
 & 0.000 & 0.198 & $0.9^{+0.6}_{-1.0}$ & $60.2^{+7.3}_{-8.5}$ & $1.8^{+0.3}_{-0.2}$ \\
 & 0.198 & 0.262 & \nodata & $32.3^{+3.8}_{-4.7}$ & $8.3^{+0.9}_{-0.8}$ \\
 & 0.262 & 0.292 & \nodata & $33.7^{+2.9}_{-3.5}$ & $37.9^{+2.4}_{-2.3}$ \\
 & 0.292 & 0.322 & \nodata & $42.3^{+2.8}_{-3.3}$ & $45.1^{+2.8}_{-2.6}$ \\
 & 0.322 & 0.352 & \nodata & $37.4^{+3.0}_{-3.6}$ & $38.3^{+2.4}_{-2.3}$ \\
 & 0.352 & 0.382 & \nodata & $34.4^{+2.9}_{-3.3}$ & $37.8^{+2.3}_{-2.2}$ \\
 & 0.465 & 0.616 & \nodata & $34.5^{+3.2}_{-4.0}$ & $16.9^{+1.3}_{-1.2}$ \\
1260 & 0.000 & 1.853 & $0.2^{+0.8}_{-0.7}$ & $37.1^{+3.3}_{-5.0}$  & $3.7 \pm 0.1$ \\
 & 0.000 & 0.523 & $1.8^{+0.5}_{-0.8}$ & $64.5^{+16.2}_{-30.1}$ & $0.4 \pm 0.1$ \\
 & 0.523 & 1.089 & \nodata & $55.0^{+6.2}_{-7.2}$ & $1.0 \pm 0.1$ \\
 & 1.089 & 1.377 & \nodata & $46.8^{+4.6}_{-5.0}$ & $1.7 \pm 0.2$ \\
 & 1.377 & 1.458 & \nodata & $36.1^{+3.1}_{-3.5}$ & $5.7^{+0.6}_{-0.5}$ \\
 & 1.458 & 1.490 & \nodata & $45.2^{+2.0}_{-2.3}$ & $48.2^{+2.6}_{-2.5}$ \\
 & 1.490 & 1.522 & \nodata & $47.8^{+2.0}_{-2.2}$ & $50.9^{+2.8}_{-2.7}$ \\
 & 1.522 & 1.554 & \nodata & $44.7^{+2.0}_{-2.2}$ & $47.8^{+2.6}_{-2.5}$ \\
 & 1.567 & 1.654 & \nodata & $43.1^{+2.9}_{-3.4}$ & $30.0^{+2.6}_{-2.4}$ \\
 & 1.783 & 1.853 & \nodata & $52.0^{+6.8}_{-7.9}$ & $3.1^{+0.6}_{-0.5}$ \\
1310 & 0.000 & 0.281 & $0.5^{+1.5}_{-1.0}$ & $33.7^{+4.9}_{-6.7}$  & $6.7 \pm 0.3$ \\
 & 0.032 & 0.059 & $4.3^{+1.3}_{-1.7}$ & $30.6^{+3.7}_{-3.8}$ & $4.8^{+0.8}_{-0.7}$ \\
 & 0.059 & 0.085 & \nodata & $45.3^{+1.7}_{-1.6}$ & $38.1^{+2.3}_{-2.1}$ \\
 & 0.085 & 0.111 & \nodata & $46.3 \pm 1.9$ & $28.6^{+2.0}_{-1.9}$ \\
 & 0.111 & 0.281 & \nodata & $40.1^{+3.7}_{-3.8}$ & $1.1 \pm 0.2$ \\
\enddata
\tablenotetext{a}{Time in seconds since T$_{start}$ as defined in Table~\ref{tab:tab2}.}
\end{deluxetable}

\section{Conclusions}
We have carried out a comprehensive search for short bursts in   data of the  INTEGRAL IBIS instrument, taken in the period  from end of February 2003 to October 2024.
The search for bursts in the light curves resulted more than 75,000 candidates. Exploiting the   \isgri\  imaging capability, we could verify the nature of each of them, which was especially important in the crowded region of Galactic Centre where multiple sources are present in the FOV. This led to a final sample of 1349 confirmed bursts from 21 different sources.

The total exposure across all sources amounted to 1136.8 Ms, with individual exposures ranging from $\sim$8 Ms for CXO0100--7211 in the Small Magellanic Cloud to nearly 68 Ms for SGRJ1745--2900, located close to the Galactic center that was extensively observed by INTEGRAL. Nine magnetars were observed for more than 50 Ms; however, only one of them, \sgros, produced a significant number of bursts. In contrast, the two next most active sources in our sample, \sgrtf\ and \onee, were observed for only $\sim$33 Ms and $\sim$27 Ms, respectively. This is not surprising as typically most of the magnetar bursts are produced during outbursts lasting weeks to months, and many sources were either not active during the \integ\ observation periods or the phases of intense activity were missed by INTEGRAL. For example, this is the case of  \sgrtf\ that underwent multiple outbursts since 2015 \citep[e.g.,][]{2020ApJ...902L..43L,2020ApJ...893..156L,2020ApJ...902L...2B,2024ApJ...965...87I,2025ApJS..276...60R},  but was observed by INTEGRAL mostly during the 2022 October activity period.

Similar to previous works, we found that the distribution of burst durations are well described by log-normal functions centred around $\sim$0.1 s and that the burst fluences follow power-law distributions with integral slopes $\beta\sim$0.7--1.
The power law distributions of burst energies suggest that a self-organized critical system might be at the bases of the sudden energy release of the bursts \citep{1996Natur.382..518C,2016SSRv..198...47A}.

We could perform a spectroscopic analysis only for the bursts with sufficient statistics (about 40\% of our sample). Most of the time averaged spectra were adequately fit with a simple power law. This resulted from  the relatively limited energy range over which the bursts were detected (which also decreased with time due to the rising lower energy threshold of ISGRI) coupled to the large statistical errors. In fact, in the bursts spectra  with good  counting statistics,  an exponential cut-off was clearly required at high energies. In these cases, the best fit peak energies \Ep\ derived with an exponentially cut-off power law model were typically in the range   $\sim$20-60 keV for \sgros\ and \sgrtf\ and slightly higher for \onee\ ($\sim$35-100 keV). 
A time resolved spectral analysis for 21  bursts with high fluence,  did not provide evidence for strong spectral evolution, with the notable exception of the \sgrtf\ burst with associated  FRB-like radio emission.

We found a negative correlation between  \Ep\  and fluence for \sgros . This confirms the reported trends of spectral softening with increasing flux \citep{2006A&A...445..313G} and fluence \citep{2001ApJ...558..228G}, that  were based only on the analysis of hardness ratios.  Negative correlations between  \Ep\  and fluence  have been seen for other magnetars, including \onee\ and \sgrtf , also with some evidence for a positive correlation at the highest fluences \citep{2012ApJ...749..122V,2015ApJS..218...11C,2020ApJ...893..156L,2011ApJ...739...87L}.
Our smaller samples of bursts from the two latter sources do not provide significant evidence for correlations,  when measurement errors are properly taken into account.

%%%%%%%%%%%%%%%%%%%%%%%%%%%%%%%%%%%%%%%%%%
%%%%%%%%%%%%%%%%%%%%%%%%%%%%%%%%%%%%%%%%%%
%%%%%%%%%%%%%%%%%%%%%%%%%%%%%%%%%%%%%%%%%%

\begin{acknowledgements}
We thank D. G{\"o}tz, C. Ferrigno and Ph. Laurent for their valuable  support and help with the  data analysis and the anonymous referee for her/his constructive comments.
We acknowledge financial support from INAF through the Magnetars Large Program Grant and from  the Italian Space Agency through the  “INTEGRAL ASI-INAF” agreement  2019-35-HH.0. 
This work is based on data obtained
with INTEGRAL, an ESA mission with instruments and science data centres funded by ESA member states, and with the participation of the Russian Federation and the USA.
\end{acknowledgements}

\facilities{INTEGRAL(IBIS/ISGRI)}
\software{astropy \citep{2013A&A...558A..33A,2018AJ....156..123A,2022ApJ...935..167A}, powerlaw \citep{2014PLoSO...985777A}, SciPy \citep{2020SciPy-NMeth}}

\bibliography{bib}{}
\bibliographystyle{aasjournalv7}

\end{document}